\documentclass[conference]{IEEEtran}
\IEEEoverridecommandlockouts
\usepackage{cite}
\usepackage{amsmath,amssymb,amsfonts}
\usepackage{algorithmic}
\usepackage{listings}
\usepackage{graphicx}
\usepackage{textcomp}
\usepackage{svg}
\usepackage{eso-pic}
\usepackage{xcolor}
\usepackage{url}
\def\BibTeX{{\rm B\kern-.05em{\sc i\kern-.025em b}\kern-.08em
    T\kern-.1667em\lower.7ex\hbox{E}\kern-.125emX}}

\usepackage{listings, xcolor}

\definecolor{verylightgray}{rgb}{.97,.97,.97}

\lstdefinelanguage{Solidity}{
	keywords=[1]{anonymous, assembly, assert, balance, break, call, callcode, case, catch, class, constant, continue, constructor, contract, debugger, default, delegatecall, delete, do, else, emit, event, experimental, export, external, false, finally, for, function, gas, if, implements, import, in, indexed, instanceof, interface, internal, is, length, library, log0, log1, log2, log3, log4, memory, modifier, new, payable, pragma, private, protected, public, pure, push, require, return, returns, revert, selfdestruct, send, solidity, storage, struct, suicide, super, switch, then, this, throw, transfer, true, try, typeof, using, value, view, while, with, addmod, ecrecover, keccak256, mulmod, ripemd160, sha256, sha3}, 
	keywordstyle=[1]\color{blue}\bfseries,
	keywords=[2]{address, bool, byte, bytes, bytes1, bytes2, bytes3, bytes4, bytes5, bytes6, bytes7, bytes8, bytes9, bytes10, bytes11, bytes12, bytes13, bytes14, bytes15, bytes16, bytes17, bytes18, bytes19, bytes20, bytes21, bytes22, bytes23, bytes24, bytes25, bytes26, bytes27, bytes28, bytes29, bytes30, bytes31, bytes32, enum, int, int8, int16, int24, int32, int40, int48, int56, int64, int72, int80, int88, int96, int104, int112, int120, int128, int136, int144, int152, int160, int168, int176, int184, int192, int200, int208, int216, int224, int232, int240, int248, int256, mapping, string, uint, uint8, uint16, uint24, uint32, uint40, uint48, uint56, uint64, uint72, uint80, uint88, uint96, uint104, uint112, uint120, uint128, uint136, uint144, uint152, uint160, uint168, uint176, uint184, uint192, uint200, uint208, uint216, uint224, uint232, uint240, uint248, uint256, var, void, ether, finney, szabo, wei, days, hours, minutes, seconds, weeks, years},	
	keywordstyle=[2]\color{teal}\bfseries,
	keywords=[3]{block, blockhash, coinbase, difficulty, gaslimit, number, timestamp, msg, data, gas, sender, sig, value, now, tx, gasprice, origin},	
	keywordstyle=[3]\color{violet}\bfseries,
	identifierstyle=\color{black},
	sensitive=false,
	comment=[l]{//},
	morecomment=[s]{/*}{*/},
	commentstyle=\color{gray}\ttfamily,
	stringstyle=\color{red}\ttfamily,
	morestring=[b]',
	morestring=[b]"
}

\begin{document}

\title{Immutable and Democratic Data in permissionless Peer-to-Peer Systems}

\author{
\IEEEauthorblockN{Maximilian Ernst Tschuchnig\\
and Dejan Radovanovic}\\
\IEEEauthorblockA{Salzburg University of Applied Sciences\\
Urstein S\"ud 1, 5412 Puch bei Hallein\\
maximilian.tschuchnig@fh-salzburg.ac.at}

\and
\IEEEauthorblockN{Eduard Hirsch\\
and Anna-Maria Oberluggauer\\
and Georg Sch\"afer}
\IEEEauthorblockA{Salzburg University of Applied Sciences\\
Urstein S\"ud 1, 5412 Puch bei Hallein\\
eduard.hirsch@fh-salzburg.ac.at}
}

\maketitle

\AddToShipoutPicture*{\small \sffamily\raisebox{1.2cm}{\hspace{1.8cm}978-1-5386-5541-2/18/\$31.00 ©2018 European Union}}

\begin{abstract}
Conventional data storage methods like SQL and NoSQL offer a huge amount of possibilities with one major disadvantage, having to use a centralized authority. This authority may be in the form of a centralized or decentralized master server or a permissioned peer-to-peer setting. This paper looks at different technologies on how to persist data without using a central authority, mainly looking at permissionless peer-to-peer networks, primarily Distributed Ledger Technologies (DLTs) and a combination of DLTs with conventional databases. Afterwards it is shown how a system like this might be implemented in two prototypes which are then evaluated against conventional databases.
\end{abstract}


\section{Introduction}

Current data storage techniques like SQL and NoSQL databases offer a huge amount of possibilities in order to handle data (e.g. ACID (Atomicity, Consistency, Isolation and Durability). But as those systems are centralized, decentralized or permissioned (in reference called hybrid p2p) peer-to-peer \cite{schollmeier2001} they all incooperate some kind of central authority. This enables those database systems to be fast and the storage to be cheap but comes with the problem of never truly being able to store data democratically\footnote{Democratically in this context means that, due to PoW, every node can participate in choosing the current state of data (assuming same calculation power, this would be virtually the same as voting)}. In these systems there is always a way to change data without asking peers for allowance (e.g. MongoDB Replication - Primary \cite{mongodb}). One way to work around this is to have every node storing the whole dataset and only allow incremental updates, if a democratic voting process from the nodes validates the changes. This voting and the resulting validation of the peers itself leads to decreased performance and high storage requirements as every node has to store the same data. These technologies are called Distributed Ledger Technologies (DLT), as every node stores its own version of the ledger, with its main representative, at the moment, the blockchain\cite{dlt}. \\

This paper investigates possibilities on how to incooperate immutability, as well as a democratically agreed on state in a blockchain. In other words, imagine data, which should be accessible to a lot of people, where not everyone is known and nobody, not even the creator of the data should be able to change it, without consent of the other participants. Scenarios like this might be in the medical sector\cite{medical}, as a tool to improve direct democracy \cite{swan2015}, where any participant can propose bills, or as a permissionless peer-to-peer certification system \cite{sproof}. In order to achieve this, two different systems are evaluated. The first system uses DLTs in order to store the data, reducing speed but also decreasing complexity\footnote{Difficulty of generating and maintaining code for given solutions} of the resulting system while the second system combines DLTs with a conventional database, increasing both complexity and scalability of the resulting system. The general contribution of this paper is to show why it is hard to have democratically agreed on data with current datastorage techniques and how this task could be solved using DLTs, specifically blockchains. \\

After the definition of important terms and setting background knowledge, this paper will evaluate an mixture of important related work like BigchainDB, ChainSQL and Sproof. Afterwards, the implementation of the different systems is explained and evaluated in comparison to several randomly chosen SQL and NoSQL databases. This evaluation will result in a discussion about advantages and problems of the explained solutions and possible improvements and future work. Finally, the paper will conclude its findings. \\

\section{Definitions and Background}

In order to accurately describe the problem of an immutable data store and possible solutions to this problem the terms immutability and blockchain as a data store have to be defined. Furthermore, as the solutions builds on the Ethereum (Eth) blockchain as well as CouchDB, these systems will also be introduced. \\

\textbf{Immutability} is defined as unchanging over time or unable to be changed\cite{immutability}. As it is impossible for anything to be unchanging over time, we define immutability, with respect to our data, to be unchangeable in a usual timeframe of about a hundred years. We also only take basic interfaces for changes into account, like the ones supposed to be used by the creators of the chosen frameworks, tools and languages and not malicious hacking. \\

\textbf{Blockchain as a Data Store} Blockchains are distributed ledgers, which basically are distributed and synchronized states on which most users agree on. The method on how this is achieved can be different (chain of blocks - Blockchain\cite{bitcoin}, Directed Acyclic Graph - Tangle\cite{iota}). Furthermore, the way on how to achieve consensus can also be different (e.g. Proof-of-Work\cite{bitcoin} or Proof-of-Stake\cite{buterin2017}). \\

\textbf{Ethereum (ETH)} is a system that uses blockchain technology in combination with a turing-complete programming language. With this language, contracts can be created to encode arbitrary state transitions on the blockchain. Therefore, Eth can be used to build fully decentralized applications (DAPs), allowing anyone in the system to write state transition functions in the form of contracts \cite{ethereum}.

This paper mainly focusses on Eth and its Smart Contract programming language solidity, but it should be noted that other blockchains like NEO\footnote{https://neo.org/}, STRATIS\footnote{https://stratisplatform.com/} or LISK\footnote{https://lisk.io/} could be used as well. Eth was used in this paper and the following implementations as it can easily be used for prototyping thanks to the web compiler Remix\footnote{https://remix.ethereum.org/} and a big community \cite{ethereum}. \\

\textbf{CouchDB} is a NoSQL database by Apache, with its focus on synchronisation. To accomplish this, it uses a specific dataformat and protocol that additionally to data, also stores the revision history of said data. On sync, the participants compare the replication histories and work out the differences of the data. This system is similar to a blockchain in terms of storing every single change (comparision blockchain: transaction) but is not immutable, as this history can be deleted without consent. Moreover, CouchDB has a centralized instance and doesn't work in a peer-to-peer setting \cite{couchdb}.

\section{Related Work}
\label{sc:Related Work}

There are several systems already dealing with the problem of peer-to-peer data storage like IPFS (although beeing a data transportation system) as well as systems combining blockchain technology with conventional databases like BigchainDB, ChainSQL and Sproof. This section will take a look a those systems, explaining them and showing their advantages and shortcomings, which the rest of this contribution will attempt to build on. \\

\textbf{IPFS} - The Interplanetary File System (IPFS) is a distributed, peer-to-peer data transportation system, similar to the internet. The main difference is that, while the internet uses location based addressing, IPFS used document based addressing. This document based addressing leads to immutability of data, as the documents are never changed. In the case of a document change, a new document is created. In order to gain trust in this peer-to-peer system, the documents get hashed, which can be checked by reconstructing the hash or merkletree from the untrusted data. IPFS is excellent at distributing data, but comes with two main disadvantages. There is no guarantee that the document, currently beeing accessed, is availiable (possible solution by using Filecoin) and there is no real option of querying  \cite{ipfs}.\\

\textbf{BigchainDB and ChainSQL} - BigchainDB as well as ChainSQL are systems intending to combine the advantages of the blockchain while not sacrificing scalability. Both systems try to achieve this by combining a blockchain with a conventional databases (BigchainDB - RethinkDB/MongoDB, ChainSQL - e.g. MySQL). This combination enables scalable queryable and immutable databases.

In \textbf{ChainSQL} each client has a full SQL database which gets updated by the blockchain on change or periodically. To add data, the client adds it to the blockchain which in turn updates all peers. This implementation comes with the problem that the blockchain has to store all the data, which cannot be deleted anymore and doesn't scale \cite{chainsql}.

The original version of \textbf{BigchainDB} ran the database on one cluster of the chosen database. This results in one illegal change being reflected to the whole cluster, removing the usefulness of the blockchain in this system \cite{bigchaindb}. This problem was identified and fixed by BigchainDB 2.0, changing from one database cluster to multiple databases per peer and using the Tendermint consensus algorithm, an alternative to PoW, to find consensus between those databases. Nonetheless, Tendermint gossip uses a centralized member list, which is strictly against our vision of a completely peer-to-peer system\cite{bigchaindb}. \\

\textbf{Sproof} is a permissionless, peer-to-peer certification system, combining the ETH blockchain with IPFS as a datastore. This system stores its data in IPFS and incorporates blockchain advantages by storing changed events and IPFS hashes of these changes in a permissionless blockchain. The described querying problem of IPFS is approached by building the data to query from the event history. This contribution builds on this idea, in combination with BigchainDb and CouchDB, in order to generalize an immutable and democratic datastore for permissionless peer-to-peer systems \cite{sproof}.

\newpage
\section{Implementation}

In this paper, two different approaches to storing data using DLTs are tested. One prototype stores all the data in the blockchain itself while another approach stores only references to the data in the blockchain and the actual data is stored in a conventional database. The first approach utilizes all advantages as well as disadvantages of the blockchain with a very simple implementation, while the second approach, for a better readability in the future called EtherCouch, minimizes the named disadvantages by adding a NoSQL database, strongly increasing the complexity of the system.

\subsection{Ethereum}
\label{sc:Ethereum}

One way of storing data in a blockchain like Eth is to put all the data in the blockchain directly or through a smart contract. To add data to a system using smart contracts, one possibility is to send a transaction to the smart contract that stores the data and a list, containing all transaction references to access the data. The first solution accomplishes this by storing data in the ethereum blockchain using smart contracts to add and keep track of all added items. For this paper the authors used a private network consisting of a single node. In order to do so the authors defined a data structure and a controller that can be viewed in their github repository\footnote{Ticket Controller Source: https://github.com/kuchenkiller\\/DatastorageInBlockchain/NodeJSPythonStoreInEth/Contracts.sol}. Since this system is built entirely using Smart Contracts on the Eth blockchain, it inherits all advantages as well as disadvantages of similar DLT platforms like NEO, STRATIS or LISK.

\subsection{EtherCouch}

The core reason for implementing EtherCouch is to combine the immutability and peer-to-peer functionality of blockchain systems with the scalability and query possibilities of conventional databases. In a way this is similar to BigchainDB and ChainSQL as the core idea is the same, but in order to circumvent the problems, described in the section related work, a new architecture and protocol was conceptualized and implemented as a prototype.

\begin{figure}[h]
\centering
\caption{EtherCouch System}
\includegraphics[width=0.50\textwidth]{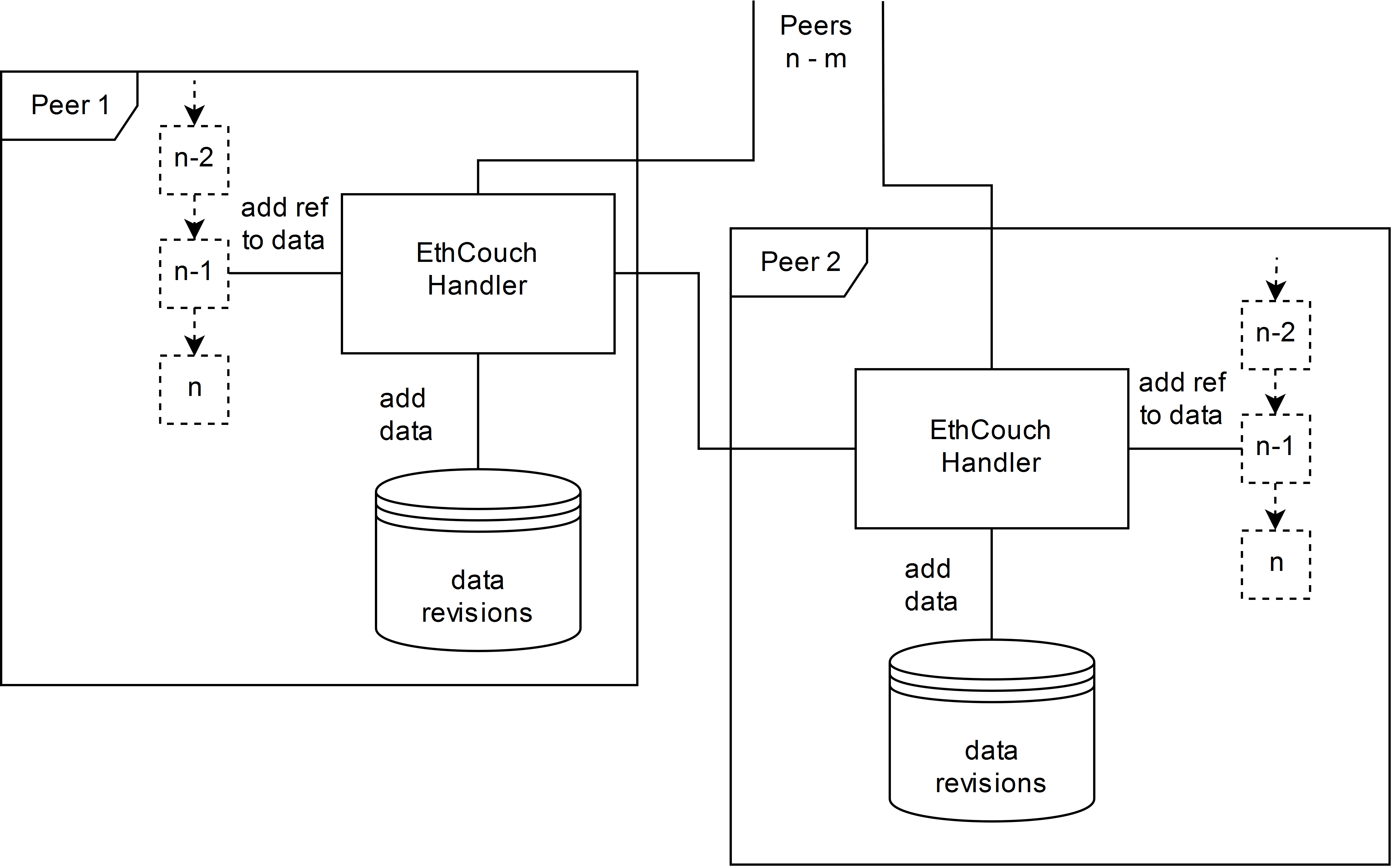}
\label{fig:etherCouchSystem}
\end{figure}

To accomplish its goals, EtherCouch, which can be viewed in figure \ref{fig:etherCouchSystem}, separates the hard to scale blockchain from the data. This is done by storing the data itself in a NoSQL database on each peer and using the blockchain (data structure can be viewed in listing \ref{lst:BlockchainData}) in order to store the hashes of the data as a reference. This results in a one-to-one relationship between the blockchain and the data. In order to include the advantages of the blockchain into a conventional database, the functionality of the blockchain must not be broken.  

Therefore, immutability and the deterministic behaviour of the blockchain have to be included in the database in order not to break the one-to-one mapping. This can be accomplished by only adding data to the database if it has been added into a block and the block has already been mined. This fixes indeterministic behaviour which could occur due to, for example, network delays.

One way to break the mapping is to have a client go offline but keep working and changing data. This would result in different, unsynchronized database states. For such case, in order to not break the determinism of the blockchain mapping, EtherCouch implements a synchronization algorithm similar to CouchDB, hence the second part of the name. In order to enable this synchronisation, EtherCouch not only stores the blockchain and the data, but also every change to the data (revisions). While this enables BASE (Basically Available, Soft-state, Eventual consistency) by comparing replication histories it comes with an obvious high additional storage cost. After reconnecting to the network, the previously offline node can generate the whole history of data changes, and not only the newest state, enabling other nodes to fill their databases in the correct way.

As illustrated in Figure \ref{fig:etherCouchSystem} the EtherCouch system is a peer-to-peer system that communicates by using on- and off-chain communication. The onchain communication enables the clients to listen to changes and tells them what hashes have been added to the blockchain. These hashes enable the clients to locate the source of the change by the editor hash, the type of change (add, edit, delete) and the data itself as the data hash. The blockchain also implements a Smart Contract to sign up clients with up to date peers and distribute the location of clients for off-chain communication purpose. Then, after locating the data, a peer can ask the source for the data off-chain and verify its integrity by hashing and comparing it to the hash in the blockchain. As the data will most likely be sent in multiple packages, verfication of the off-chain distributed data can be done by using merkle trees, similar to how the Interplanetary File System (IPFS) handles data separated into multiple packages \cite{ipfs}.

\begin{figure}[h]
\centering
\caption{EtherCouchAddData}
\includegraphics[width=0.48\textwidth]{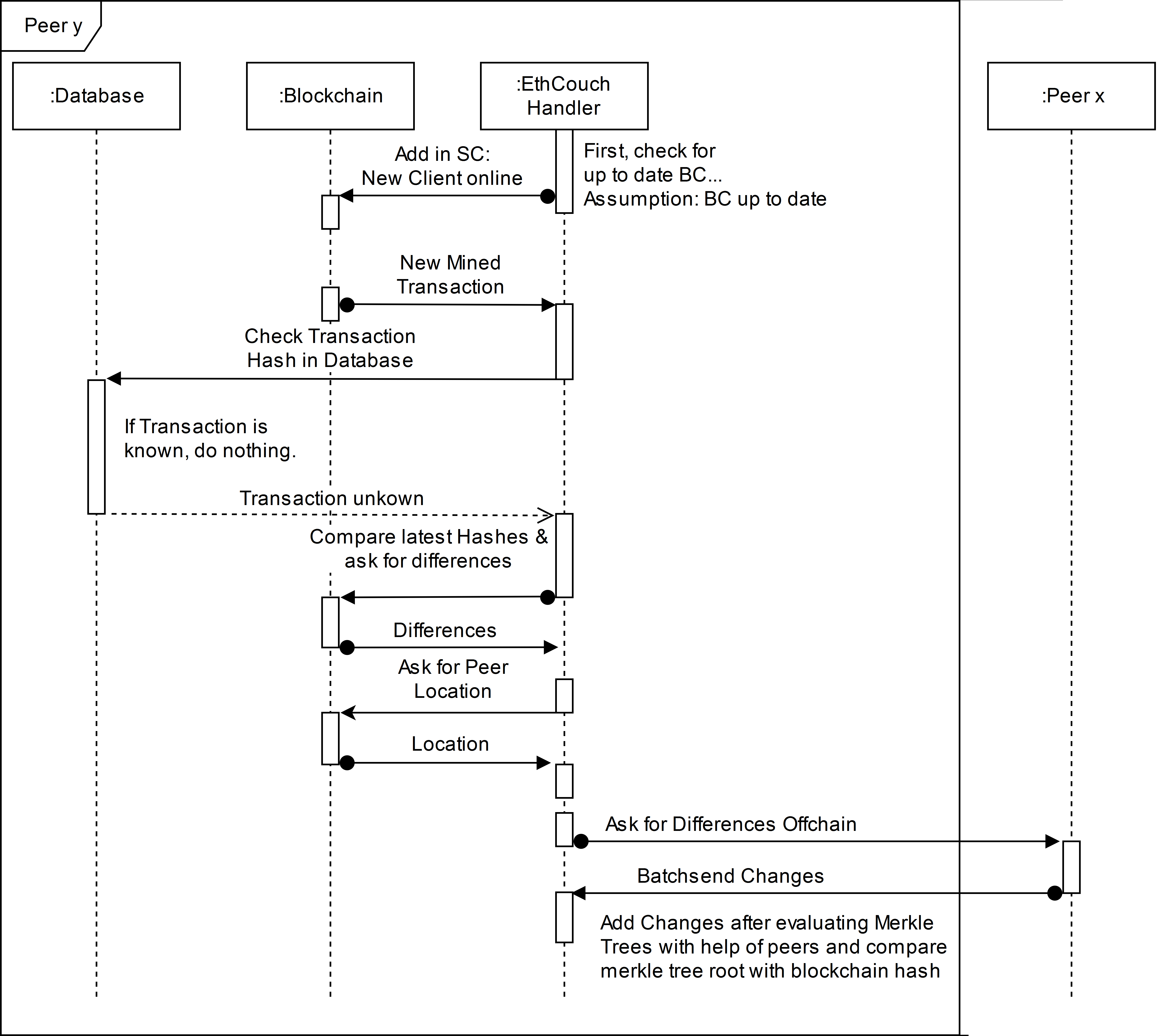}
\label{fig:EtherCouchAddData}
\end{figure}

An additional TopicId field in the blockchain data can be used in order to only listen to changes, relevant to peers. This TopicID can also be used by the replication source in order to check if the off-chain peer is allowed to get the data which it is currently trying to access.

The basic protocol enables adding, editing and deletion of data. The view of data is possible without accessing the blockchain, as each client has access to its own database, vastly increasing read speed. This leads to a huge privacy risk that can be approached by applying public private key cryptogaphy to secure data readability\cite{pbpk}. With the current findings on quantum computing in mind, using post-quantum cryptography should be taken into account for future expansions on this paper \cite{bernstein2011}.

\begin{lstlisting}[language=Solidity,caption={Blockchain Datastruct}, captionpos=b, label={lst:BlockchainData}]
struct DbFunction {
	bytes32 task;
	bytes32 data_hash;
	bytes32 editor_hash;
	bytes32 topic_id;
	uint256 sequence_id;
}
\end{lstlisting}

In order to add data to the database, a user has to generate the data and hash it. Then the hash has to be pushed into the data
storage Smart Contract (similar to section 4 Ethereum) in combination with the already explained metadata (listing \ref{lst:BlockchainData}). After the node adds the change to the blockchain, it will tell the Location Contract that it is up-to-date with the newest blockchain hash (the recently added hash). If the new transaction gets put into a block, the miners will recognize this on mining and will look for the new transaction. In order to propagate the actuall data after mining, the data will be broadcast offchain. Alternatively the data is already added and confirmed so the peers can ask the Location Contract (listing \ref{lst:LocationContract}) for the location of the data\footnote{At least one peer should always be online, propagation the newest data (original creator). Theoretical, there is the possibility that the creator goes offline without propagating the data off-chain, resulting in the same problem IPFS has, as described in the related work section} and query the data off-chain in order to add it to their own database.

\begin{lstlisting}[language=Solidity,caption={Location Contract}, captionpos=b, label={lst:LocationContract}]
contract LocationContract {
    struct peer_location { bytes32 editor_hash; bytes32 location; }
    peer_location[100] up_to_date_peers;
    peer_location[100] all_peers;
    
    function get_peer_location(bytes32 peer) public returns(bytes32) {
        for (uint8 i = 0; i< 100; i ++)
            if (peer == all_peers[i].editor_hash)
                return all_peers[i].location;
    }
    
    function get_up_to_date_peer_location() public returns(bytes32) {
        return up_to_date_peers[0].location;
    }
}
\end{lstlisting}

This system never actually changes data, it only makes old data obsolete by adding the changed data and making it the currently active data (the old data becomes the latest revision and the new data becomes the actual data). Therefore, an edit is almost the same as an add. The only difference is, that after querying the changed data the new data revision id is incremented by one, effectively making it the active revision. 

As a blockchain is immutable, deleting data should be impossible, which is a problem regarding the new GDPR (General Data Protection Regulation) in Europe. EtherCouch enables us to delete data again while also utilising the immutability of the blockchain. In order to delete data in EtherCouch, a delete request is entered into the blockchain. This request will be propagated through the blockchain and tell nodes to delete the respective data. Although this will not delete the references in the blockchain, the actual data will be deleted. This seems to be a huge security issue, as nobody can be forced to delete data, but conventional systems suffer the same problem. If a database node is untrustworthy, this deletion might not occur (extreme example would be printing every transaction on paper). One solution to this problem, independent of this contribution is to only allow the data to trustworthy peers. This approach will later be further discussed but as of right now, this problem remains unsolved by this contribution.\\

The described system has also been partially implemented as a prototype at the time of writing. The implementation was done by using a WebApp as a client and user interface and the ethereum blockchain as the decentralized application platform. Although it may not be able to build a peer-to-peer system with only this approach it was used in order to simulate a mobile client, using EtherCouch in a slow environment. For the client IndexDB\footnote{\url{https://www.w3.org/TR/IndexedDB-2/}} was used as the basic database and revisions database. The reason for this paper was not to show a finished product but rather show how an immutable database could be created. Therefore, a full implementation was not achieved since it is not needed to show the general system and time critical tasks, and working with the blockchain, can still be tested. \\

A further improvement which is only partially implemented yet is the ability of using data filters. Since an editor hash is added into the blockchain transactions (listing \ref{lst:BlockchainData}), data distributors can check if the replicating peer is actually allowed to replicate the data. This would remove storage requirements for peers but increase complexity for the Location Contract (listing \ref{lst:LocationContract}) since not every up to date peer now has all data stored and peers would need hash lists of stored data.

\section{Evaluation}
In order to evaluate the prototypes and check how they scale, a test case was created and evaluated with conventional databases as a baseline. As the addition of new data into the blockchain is expected to be the slowest part of the whole system, this scenario was chosen for our evaluation. The following evaluation measures the time for creating $10^1 - 10^6$ datasets in the form of maintenance tickets. As a baseline to the prototypes the conventional databases Microsoft SQL Server Express, MongoDB and the synchronisation optimized database CouchDB (in sync with a client side PouchDB) was used and the reported values are the mean values of 5 runs of the experiment.

\begin{figure*}[h]
\centering
\caption{Evaluation Prototypes - Conventional Databases}
\includegraphics[width=1\textwidth]{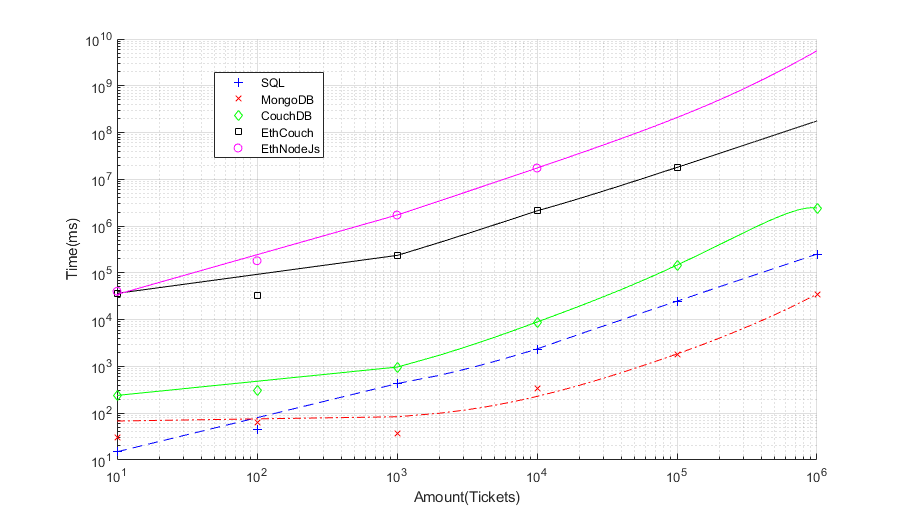}
\label{fig:evaluierung}
\end{figure*}

As it can be seen in figure \ref{fig:evaluierung} the conventional approaches are less time consuming, with respect to increasing execution times, than the blockchain approaches. Storing the data only in the Eth blockchain yields to the worst result, especially if the ticket amount increases. Since EtherCouch does not store the data itself but only the data hashes it scales a lot better with an increase in data amount and amount of data per hash. The data described in figure \ref{fig:evaluierung} was obtained in the following environment: Windows 10 Enterprise as a operating system, geth version 1.8.16 as the Eth platform and Mozilla Firefox version 61.0 as well as nodejs version 8.12.0 to run the clients.\\

In conclusion, it can be said that the blockchain approaches scale worse than the tested conventional systems, but definitely have advantages which will be summarized in the next section. EthCouch scales better than the pure Blockchain solutions and, with a high amount of tickets, getting close to CouchDB\footnote{This scalability improvement, compared to storing data only in blockchains increases as the individual datasize (in our case ticket size) increases, since the hash always stays the same size}. That means that EtherCouch might be used in non time critical applications with the need for democratically agreed on data.

\section{Advantages and Problems}

This section will talk about the main advantages of both proposed solutions and their main problems. We acknowledge that no one persisting method fits all needs, therefore we will take the opportunity to talk about the main advantages and disadvantages of the proposed systems.\\

\textbf{Advantages} - Immutability as well as a democratically agreed on state are by far the most important advantages of a blockchain based system as there are hardly any other possibilities on how to achieve both properties. Another advantage of storing data in a blockchain is that synchronization is comes by design, since the system has a democratically agreed on state. EtherCouch, which stores data in a conventional database and only saves the references to this data in a blockchain by design, stores as little data as needed in the blockchain itself. Therefore, EtherCouch is scalable as well as query-able in contrast to other blockchain systems. The following enumeration gives an overview of the advantages of both proposed solutions.

\quad\textit{All}
\begin{itemize}
\item Immutable (if peers keep to the protocol)
\item Democratic state
\item Easy synchronization
\end{itemize}
\quad\textit{EtherCouch}
\begin{itemize}
\item Scalable
\item Querying depending on used database
\item Blockchain storage minimal \\
\end{itemize}

\textbf{Problems} - Storing data fully or partially in a distributed ledger also comes at a high cost. First of all, as a distributed ledger is a distributed data store where every full node has to store the complete database, a lot of data is stored multiple times. Furthermore, the basic consensus algorithm for permissionless blockchains, PoW, can have negative side-effects as, in most PoW systems, mining is rewarded \cite{bitcoin}. This leads to a battle of miners to get more calculation power than the opposition which has two negative effects. The first being the unnecessary usage of energy, as this increased difficulty does not help the blockchain in any way. Secondly, this behavior leads to miners assembling into mining pools in order become faster miners. Taken to the extreme, these mining pools destroy the democratic character of PoW and the security that comes with it (51\% attacks)\cite{ethereum}. Another problem comes from the basic architecture of distributed ledgers (Blockchain and DAGs). As those systems are basically lists/graphs, they do not have any query capabilities by design (other than walking through the data structure), making querying slow and requiring extra measures. As EthCouch implements a conventional database as its main data store, it is not affected by this problem. A second problem is the speed of adding data as new data has to be approved by all peers. As Distributed Ledger Technologies (DLT) are still pretty new, and a lot of systems are still prototypes without any production standard there are little to no implementation rules or design patterns. Furthermore, Libraries and Frameworks still tend to change a lot, introducing breaking changes along the way and changing interfaces periodically. This increases the difficulty to work with DLTs and definitely should be taken into account when starting a DLT project. The following enumeration gives an overview of the problems of both proposed solutions.

\newpage
\textit{All}
\begin{itemize}
\item PoW has high energy demands
\item Low speed
\item Little standards
\end{itemize}
\quad\textit{Only Blockchain}
\begin{itemize}
\item High storage requirement in blockchain
\item No Querying \\
\end{itemize}

\section{Conclusion and Future Work}

In conclusion, conventional data storage methods are fast and the preferred option for large amounts of data, but for any use case where the speed of saving data does not matter or immutability as well as a democratically agreed on state plays a main role, both prototypes as well as the systems described in section \ref{sc:Related Work} should be taken into account. Fully storing a lot of data in a blockchain is simple but doesn't scale, while a system like EthCouch scales, with the price of greatly increasing complexity\footnote{Difficulty of generating and maintaining code for given solutions}.

In the future, encryption can be implemented by a public-private-key infrastructure, in order to store data without being publicly visible. It is also possible to add a two part validation system, enabling data to be either validated or pending. When adding data, a node now additionally adds the new data as \textit{validation=pending} into a messaging queue like Kafka. This enables other nodes, which take the risk of working with pending data, to listen to this queue and vastly increase their speed. 

Another improvement would be to create a specialized blockchain for EthCouch in which a useful PoW algorithm is implemented, without transaction fees. In order to reward mining in such a system other methods of payments, like free cloud storage, which could again be hosted by peers, have to be found. Furthermore, the addition of data filters reduces the amount of data and revisions each peer has to store. In future work, these possibilities will be investigated by, for example, updating the Location Contract to enable filtered peer-to-peer replication. 


\bibliography{bibliography}
\bibliographystyle{plain}

\end{document}